\documentclass[conference]{IEEEtran}
\IEEEoverridecommandlockouts
\usepackage[linesnumbered,ruled,vlined]{algorithm2e}
\usepackage{cite}
\usepackage{amsmath,amssymb,amsfonts}
\usepackage{graphicx}
\usepackage{url}
\usepackage{textcomp}
\usepackage{xcolor}
\def\BibTeX{{\rm B\kern-.05em{\sc i\kern-.025em b}\kern-.08em
    T\kern-.1667em\lower.7ex\hbox{E}\kern-.125emX}}

\begin{document}

\title{Empowering Healthcare through Privacy-Preserving MRI Analysis}

\author{
    \IEEEauthorblockN{
        Al Amin$^{1}$, Kamrul Hasan$^{1}$, Saleh Zein-Sabatto$^{1}$, Deo Chimba$^{1}$, Liang Hong$^{1}$, Imtiaz Ahmed$^{2}$, Tariqul Islam$^{3}$
    }
    \IEEEauthorblockA{
        $^{1}$ Tennessee State University, Nashville, TN, USA \\
        $^{2}$Howard University, Washington, DC, USA \\
        $^{3}$ Syracuse University, Syracuse, NY, USA \\
        Email: $\lbrace$\textit{aamin2, mhasan1, mzein, dchimba, lhong}$\rbrace$@tnstate.edu,  $\lbrace$\textit{imtiaz.ahmed}$\rbrace$ @howard.edu, $\lbrace$\textit{mtislam}$\rbrace$ @syr.edu
    }
}
\maketitle

\begin{abstract}
In the healthcare domain, Magnetic Resonance Imaging (MRI) assumes a pivotal role, as it employs Artificial Intelligence (AI) and Machine Learning (ML) methodologies to extract invaluable insights from imaging data. Nonetheless, the imperative need for patient privacy poses significant challenges when collecting data from diverse healthcare sources. Consequently, the Deep Learning (DL) communities occasionally face difficulties detecting rare features. In this research endeavor, we introduce the Ensemble-Based Federated Learning (EBFL) Framework, an innovative solution tailored to address this challenge. The EBFL framework deviates from the conventional approach by emphasizing model features over sharing sensitive patient data. This unique methodology fosters a collaborative and privacy-conscious environment for healthcare institutions, empowering them to harness the capabilities of a centralized server for model refinement while upholding the utmost data privacy standards. Conversely, a robust ensemble architecture boasts potent feature extraction capabilities, distinguishing itself from a single DL model. This quality makes it remarkably dependable for MRI analysis. By harnessing our groundbreaking EBFL methodology, we have achieved remarkable precision in the classification of brain tumors, including glioma, meningioma, pituitary, and non-tumor instances, attaining a precision rate of 94\% for the Global model and an impressive 96\% for the Ensemble model. Our models underwent rigorous evaluation using conventional performance metrics such as Accuracy, Precision, Recall, and F1 Score. Integrating DL within the Federated Learning (FL) framework has yielded a methodology that offers precise and dependable diagnostics for detecting brain tumors.
\end{abstract}

\section{Introduction}
Identifying a brain tumor is a critical and life-altering event for patients. Globally, approximately 190,000 cases of both primary and secondary brain tumors are diagnosed each year, impacting individuals of all age groups \cite{arif2022brain}. Physicians extensively utilize magnetic resonance (MR) imaging to ascertain the presence and characteristics of tumors \cite{ari2018deep,ranjbarzadeh2023brain,amin2024explainable}. Therefore, researchers diligently concentrate on AI, ML, and DL frameworks to extract meaningful features from MR images \cite{ari2018deep,maqsood2022multi,saeedi2023mri,wu2023aggn}. Nonetheless, achieving enhanced performance with DL architecture necessitates a substantial volume of data \cite{garcea2023data,kebaili2023deep,biswas2023data,alomar2023data,amin2022industrial}. The exchange of patient data among various healthcare providers is complicated by the need to protect patient confidentiality \cite{fezai2023deep,akbar2023beware,weigand2023accelerated}. This situation makes it difficult to gather more data from different hospitals \cite{kaissis2020secure,willemink2020preparing}. 
In this research, our focus centers on the precise identification of various brain tumor types, including glioma, meningioma, pituitary tumors, and non-tumorous conditions, through the utilization of MR images. To achieve this objective while respecting patient data privacy, we introduce the novel EBFL framework. This innovative approach enables individual hospitals to train their data with AI tools, sharing their feature-extracted trained models at a central server aggregation point without needing patient data exchange. The main server model leverages potent feature extraction techniques, amalgamating insights from diverse hospitals. Additionally, we employ an ensemble model, incorporating multiple DL architectures, to counter model overfitting and harness various feature extraction methods inherent to distinct DL architectures. This comprehensive strategy enhances our diagnostic algorithm's reliability and robustness, surpassing a single algorithm's capabilities.
\begin{itemize}
 \item First, the research introduces a robust Federated Learning (FL) model incorporating a maximum voting classifier (Ensemble techniques) based on DL architecture designed for a privacy-preserving hybrid paradigm. This integration is intended to enhance diagnostic reliability and precision, providing healthcare practitioners with improved tools for detecting brain tumors efficiently.

\item Second, our research introduces the EBFL Framework. This novel approach upholds patient data confidentiality while enabling multiple healthcare institutions to benefit from a centralized diagnostic model. This framework allows for synthesizing a global model's intelligence, with participating entities contributing to and deriving insights from the model without exchanging sensitive data.  
\item Finally, we developed a robust ensemble mode for training local data by combining multiple Transfer Learning architectures MobileNetV2, VGG16, and VGG19.

\end{itemize}

\section{Related Works}
DL architectures are crucial in healthcare research, enhancing MRI analysis for applications such as COVID-19 classification \cite{goyal2023detection,nasir2023multi,ullah2023densely,al2022adpt}, spinal cord disorders \cite{sollmann2021mri,jentzsch2021spinal,santiago2020role}, cancer detection and staging \cite{de2020deep,wu2020development,abhisheka2023comprehensive}, musculoskeletal disorders \cite{fritz2023radiomics,kamiya2020deep}, abdominal and pelvic disorders \cite{hemke2020deep}, and brain and neurological disorders \cite{zhang2020survey,yamanakkanavar2020mri,1}. However, Brain tumor recognition is a challenging task. A notable example of advancements in brain tumor classification, specifically targeting glioma, meningioma, and pituitary tumors, is the development of DeepTumorNet, a hybrid DL model. This model innovatively adapts the GoogleNet architecture, a CNN variant. In creating DeepTumorNet, the original architecture was modified by removing the last five layers of GoogleNet and integrating fifteen new layers in their place. This adaptation has led to remarkable outcomes. However, it's important to note the limitations of this approach. DeepTumorNet requires a substantial amount of data for training and is characterized by its computational complexity, which could pose challenges in practical healthcare applications due to patient privacy\cite{raza2022hybrid}. Another research group applies a deep CNN for brain tumor classification from MR images. A notable limitation is the small size of the dataset used for training, which can lead to overfitting and reduced generalizability of the model. Additionally, the high variance in MR images due to different scanning protocols and machines poses a challenge for accurate classification. The research suggests the need for larger, standardized datasets and improved models for better diagnostic accuracy in medical imaging \cite{ayadi2021deep}. A new set of researchers applied a two-channel deep neural network (DNN) model for brain tumor classification, utilizing pre-trained CNNs like InceptionResNetV2 and Xception for feature extraction. The model emphasized tumor-affected regions using an attention mechanism and achieved strong generalization on benchmark datasets. A limitation of this research is its focus on only two specific pre-trained models. Future work plans to explore additional pre-training models and develop a new method for abstracting features from convolution blocks \cite{bodapati2021joint}. A diverse research team developed a method for brain tumor classification that combines an ensemble of deep features with machine learning classifiers. This approach utilizes transfer learning with various pre-trained DCNN to extract features from brain MR images, which are then evaluated by multiple classifiers, including a notably effective support vector machine (SVM) with radial basis function (RBF) kernel, particularly for large datasets \cite{kang2021mri}. Another consortium of researchers used a methodology involving data acquisition, preprocessing, feature extraction using texture features, feature selection, and classification with a Support Vector Machine (SVM) ensemble model. The limitation was the focus on intermediate-stage tumor data, which might impact the model's effectiveness in early-stage tumor classification. This is an area for future improvement, along with enhancing the robustness of the SVM learner and voting algorithm \cite{shafi2021classification}. An alternative scientific group mentions various applications of FL in medical imaging, like MRI reconstruction with decentralized training of generative image priors, distinguishing healthy and cancerous brain tissues, explicitly highlighting its use in brain tumor segmentation and classification. The limitations of FL in this research area, such as communication challenges, data bias, and the need for performance improvement strategies, are also addressed \cite{nazir2023federated}. In this research, we have successfully addressed the limitations of existing FL applications in brain tumor classification by developing an innovative ensemble model. This model synergistically combines three DL architectures: MobileNetV2, VGG16, and VGG19. Each architecture employs distinct feature extraction techniques, which are particularly adept at handling small datasets with rare features. This approach enhances performance and effectively mitigates the risk of model overfitting. Furthermore, our implementation of federated learning plays a crucial role in preserving patient privacy, as it focuses on feature analysis without necessitating the sharing of raw data. Significantly, we have also introduced a novel framework that developed communication among healthcare industries.

\section{Methodology}
\begin{figure*}[h]
    \centering
    \includegraphics[height=7cm, width=\linewidth]{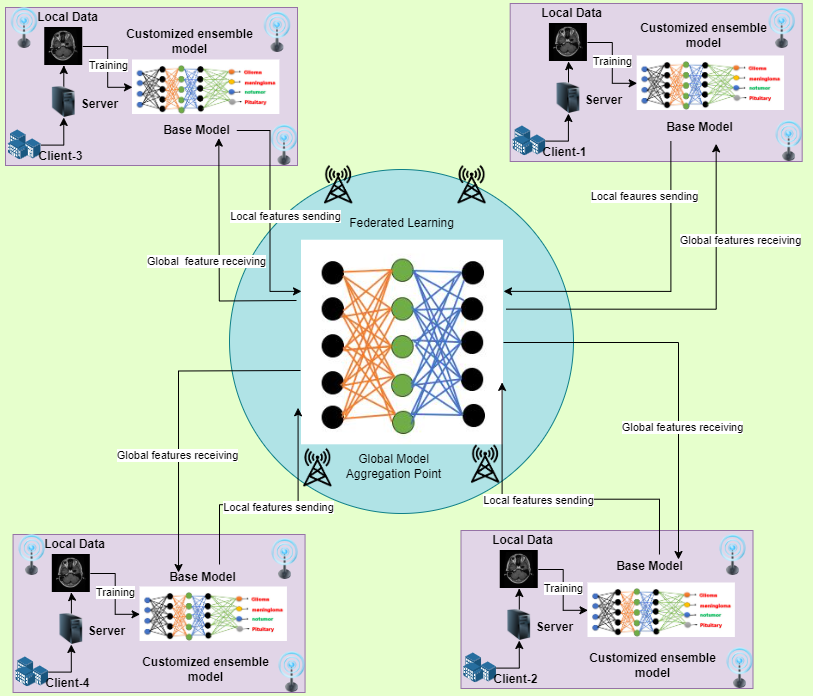}
    \caption{Ensemble Based Federated Learning (EBFL) framework for analyzing MR Images}
    \label{fig:my_label1}
    \vspace{-4mm}
\end{figure*}

The EBFL framework, as shown in Figure \ref{fig:my_label1}, integrates multiple DL algorithms for MRI image analysis, using majority voting and feature sharing at the aggregation point to enhance global model robustness and ensure data privacy, effectively addressing the challenges of data scarcity.

\subsection{EBFL based MR images System Description}

The proposed EBFL framework for brain tumor MRI analysis is structured into three sequential phases, as illustrated in Figure \ref{fig:my_label1}. Initially, the framework decentralizes the comprehensive brain tumor MRI dataset across four distinct clients. Subsequently, each client independently trains an ensemble of models on their respective server, leveraging the strength of combined algorithms for robust feature extraction. In the culmination phase, clients contribute their learned features to a central aggregation point, culminating in constructing a Federated Global Model. By virtue of collaborative learning, this model is anticipated to surpass the predictive capabilities of any single client's model, thereby enhancing the efficacy of real-time image classification.

 Data Decentralization and Classification: In the initial phase, the framework undertakes the task of data decentralization, where MRI brain tumor data, encompassing four distinct classes—Gliomas, Meningiomas, Pituitary tumors, and non-tumorous images—are distributed among four clients (num\_clients=4). This stratification ensures that each client holds a representative subset of the entire data spectrum, facilitating the diversity and richness required for a comprehensive learning process.

Ensemble Model-Based MRI Analysis on a Local Server:
The second phase is dedicated to the local training of models on each client's server. Where an ensemble classification approach utilizing $M$ distinct classifiers, denoted as $h_m(X)$, for dataset $X$. The final classification, $C_X$, is determined by a majority vote, or mode, of the outcomes from all classifiers:

\begin{equation}
C_X = \text{mode} { h_1(X), \dots, h_M(X) }.
\end{equation}

Weights $w_j$ are assigned to each classifier to influence their contribution to the final decision, leading to a weighted majority vote mechanism:

\begin{equation}
C_X = \arg\max_i \sum_{j=1}^M w_j \mathbb{I}(h_j(X) = i),
\end{equation}

where $\mathbb{I}$ is the indicator function. This method effectively combines multiple predictions, leveraging the strengths of diverse classifiers. The ensemble approach's ability to mitigate the risk of overfitting, thereby ensuring robust feature extraction and model generalization.

Federated Aggregation and Global Model Development: In the final phase, the individually trained models from each client undergo a process of federated aggregation. Here, the unique feature extraction insights gleaned from each client's ensemble model are shared with a central aggregation point. This collaborative effort forms a Federated Global Model, which amalgamates distributed intelligence into a singular, potent model. Subsequently, this enhanced global model is redistributed to each client-server, enabling the prediction of real-time images with augmented accuracy and reliability.

 Algorithm \ref{alg:EEFLF} introduces a groundbreaking approach to medical image analysis by combining multiple DL models with FL, enhancing diagnostic accuracy while preserving privacy and integrity.

\begin{algorithm}[ht]
\scriptsize
\caption{EBFL Framework for MRI Analysis}\label{alg:EEFLF}
\DontPrintSemicolon 
\KwData{MRI\_Brain\_tumor\_data, num\_clients=4}
\KwResult{Enhanced prediction accuracy for real-time images}

\SetKwFunction{FMain}{Main}
\SetKwProg{Pf}{Function}{:}{}
\SetKwFunction{FTrain}{TrainClientModel}
\SetKwFunction{FAggregate}{AggregateModels}
\SetKwFunction{FPredict}{MakePrediction}

\Pf{\FMain{}}{
    \For{client\_id \KwTo num\_clients}{
        local\_data $\gets$ DistributeData(MRI\_Brain\_tumor\_data, client\_id)\;
        local\_model $\gets$ \FTrain{local\_data, client\_id}\;
        SendToServer(local\_model)\;
    }
    global\_model $\gets$ \FAggregate{}\;
    \For{client\_id \KwTo num\_clients}{
        \FPredict{global\_model, client\_id}\;
    }
}
\SetKwProg{Fn}{Function}{:}{}
\Fn{\FTrain{data, client\_id}}{
    architectures $\gets$ [MobileNetV2, VGG16, VGG19]\;
    ensemble\_predictions $\gets$ [ ]\;
    \For{architecture in architectures}{
        model $\gets$ InitializeModel(architecture)\;
        trained\_model $\gets$ TrainModel(model, data)\;
        ensemble\_predictions.append(trained\_model)\;
    }
    \Return CombineModels(ensemble\_predictions, method='voting')\;
}
\Fn{\FAggregate{}}{
    aggregated\_models $\gets$ ReceiveFromClients()\;
    weighted\_models $\gets$ CalculateWeights(aggregated\_models)\;
    global\_model $\gets$ WeightedSum(weighted\_models)\;
    \Return global\_model\;
}

\Fn{\FPredict{model, client\_id}}{
    test\_data $\gets$ LoadClientTestData(client\_id)\;
    prediction $\gets$ model.Predict(test\_data)\;
    \Return prediction\;
}

\end{algorithm}

\subsection{Mathematical Formulation for the Proposed Maximum Voting Classifier Techniques incorporating federated learning}

Given a set of \( N \) classes for brain tumor classification, let \( \mathcal{C} = \{c_1, c_2, \ldots, c_N\} \) represent the class labels for glioma, meningioma, pituitary, and no-tumor. For a dataset \( X \) with \( m \) MRI images, \( X = \{x_1, x_2, \ldots, x_m\} \), and a set of \( K \) base deep learning models \( \mathcal{M} = \{M_1, M_2, \ldots, M_K\} \), the ensemble-enhanced federated learning framework operates as follows:

Data Distribution: The dataset \( X \) is partitioned into \( P \) subsets corresponding to the number of clients, such that \( X_p \subset X \) for \( p = 1, \ldots, P \), where \( P \) is the total number of clients.

Local Model Training: Each client \( p \) trains a local ensemble model \( E_p \) using its subset of data \( X_p \), where \( E_p \) is a combination of models \( \mathcal{M} \) trained on \( X_p \). The ensemble model at client \( p \) is defined as \cite{17}:
\vspace{-5mm} 

\begin{equation}
E_p(x) = \text{mode} \{ M_1(x), M_2(x), \ldots, M_K(x) \} \quad \text{for} \quad x \in X_p
\end{equation}
\vspace{-2mm} 

Global Model Aggregation: After training, clients contribute their model's knowledge to a central server. The server aggregates these updates to form a global model \( G \), which is then distributed back to the clients for prediction on new data.

Prediction with the Global Model: The global model \( G \) is used to predict the class label \( y \) for a new MRI image \( x \) as follows:
\begin{equation}
y = G(x) = \text{mode} \{ E_1(x), E_2(x), \ldots, E_P(x) \}
\end{equation}

The ensemble model's prediction \( y \) for an input instance \( x \) is the class label that receives the majority vote from the predictions of the local ensemble models \( E_p \).

For the loss function, the multiclass cross-entropy loss for a dataset with \( m \) samples and \( N \) classes is used to optimize the ensemble model and is defined as \cite{17}:
\begin{equation}
L = -\frac{1}{m} \sum_{i=1}^{m} \sum_{j=1}^{N} y_{ij} \log(\hat{y}_{ij})
\end{equation}
where \( y_{ij} \) is the binary indicator (0 or 1) if class label \( j \) is the correct classification for observation \( i \), and \( \hat{y}_{ij} \) is the predicted probability of observation \( i \) being of class \( j \).

Algorithm 2 captures the essence of using multiple classifiers in a federated learning setting to address the challenge of brain tumor classification, emphasizing preserving data privacy and utilizing the strengths of ensemble learning.

\begin{algorithm}[ht]
\scriptsize
\caption{EBFL Framework for MRI Analysis}\label{alg:EEFLF}
\DontPrintSemicolon 
\KwData{MRI\_Brain\_tumor\_data, num\_clients=4}
\KwResult{Enhanced prediction accuracy for real-time images}

\SetKwFunction{FMain}{Main}
\SetKwProg{Pf}{Function}{:}{}
\SetKwFunction{FTrain}{TrainClientModel}
\SetKwFunction{FAggregate}{AggregateModels}
\SetKwFunction{FPredict}{MakePrediction}

\Pf{\FMain{}}{
    \For{client\_id \KwTo num\_clients}{
        local\_data $\gets$ DistributeData(MRI\_Brain\_tumor\_data, client\_id)\;
        local\_model $\gets$ \FTrain{local\_data, client\_id}\;
        SendToServer(local\_model)\;
    }
    global\_model $\gets$ \FAggregate{}\;
    \For{client\_id \KwTo num\_clients}{
        \FPredict{global\_model, client\_id}\;
    }
}

\SetKwProg{Fn}{Function}{:}{}
\Fn{\FTrain{data, client\_id}}{
    architectures $\gets$ [MobileNetV2, VGG16, VGG19]\;
    ensemble\_predictions $\gets$ [ ]\;
    \For{architecture in architectures}{
        model $\gets$ InitializeModel(architecture)\;
        trained\_model $\gets$ TrainModel(model, data)\;
        ensemble\_predictions.append(trained\_model)\;
    }
    \Return CombineModels(ensemble\_predictions, method='voting')\;
}

\Fn{\FAggregate{}}{
    aggregated\_models $\gets$ ReceiveFromClients()\;
    weighted\_models $\gets$ CalculateWeights(aggregated\_models)\;
    global\_model $\gets$ WeightedSum(weighted\_models)\;
    \Return global\_model\;
}

\Fn{\FPredict{model, client\_id}}{
    test\_data $\gets$ LoadClientTestData(client\_id)\;
    prediction $\gets$ model.Predict(test\_data)\;
    \Return prediction\;
}

\end{algorithm}

\subsection{Multi-Class Brain Tumor Classification Using Federated Learning mathematical explanation}

Each client $C_i$ trains a local model $M_i$ on their own dataset $D_i$. The objective is to minimize the local loss function $L_i$, which is a function of the model parameters $\theta_i$ and the data $D_i$:
\begin{equation}
L_i(\theta_i) = \frac{1}{|D_i|} \sum_{(x,y) \in D_i} l(M_i(x;\theta_i), y)
\end{equation}
where $l$ is a loss function like cross-entropy for classification tasks, $x$ is the input data, and $y$ is the label.

\subsection*{Local Model Evaluation}
After training, each client evaluates the performance of their model on a validation set to calculate metrics such as accuracy, precision, recall, and F1-score.

\subsection*{Model Aggregation (Federated Averaging)}
The server aggregates the parameters $\theta_i$ from all clients to update the global model $M_g$ using federated averaging. The new global parameters $\theta_g$ are computed as the weighted average of local parameters:
\begin{equation}
\theta_g = \frac{\sum_{i=1}^{N} w_i \theta_i}{\sum_{i=1}^{N} w_i}
\end{equation}
where $w_i$ is the weight for client $i$, often chosen as the number of samples in $D_i$, and $N$ is the total number of clients.

\section{ Experimental Analysis}
\subsection{Dataset Description \& experimental result}

Our research employs a comprehensive dataset of 6,771 brain MRI scans sourced from the Kaggle data repository \cite{2} to classify brain tumors. This dataset encompasses four distinct categories: glioma (1,621 images), characterized by aggressive tumors arising from the glial cells; meningioma (1,645 images), typically slower-growing tumors originating from the meninges; pituitary tumors (1,775 images), often benign growths in the pituitary gland; and non-tumorous scans (2,000 images), serving as a control group. The dataset is split into 80\% for training and 20\% for testing, allowing for comprehensive analysis and validation of the classification performance. Table 1 shows the full classifier training and validation results of five different approaches.

\begin{table}[h]
\label{tab1}
\caption{Comprehensive Assessment of Model Performance}
\resizebox{\linewidth}{!}{
\begin{tabular}{|l|l|l|l|l|l|l|l|}
\hline
Algorithms                                               & \begin{tabular}[c]{@{}l@{}}Precision\\  (\%)\end{tabular} & \begin{tabular}[c]{@{}l@{}}Recall\\  (\%)\end{tabular} & \begin{tabular}[c]{@{}l@{}}F1- \\ Score \\ (\%)\end{tabular} & \begin{tabular}[c]{@{}l@{}}Training \\ Accuracy\\    (\%)\end{tabular} & \begin{tabular}[c]{@{}l@{}}Training\\  loss (\%)\end{tabular} & \begin{tabular}[c]{@{}l@{}}Validation\\ Accuracy\\    (\%)\end{tabular} & \begin{tabular}[c]{@{}l@{}}Validation\\ loss   \\ (\%)\end{tabular} \\ \hline
Global Model (FL)                                                    & 94                                                       & 91                                                     & 92                                                           & 95                                                                     & 0.005                                                         & 94                                                                      & 0.06                                                                \\ \hline

MobileNetV2                                                & 94.19                                                        & 92                                                     & 94                                                           & 94                                                                     & 0.06                                                           & 96                                                                      & 0.06                                                                \\ \hline
VGG16                                              & 90.62                                                        & 91                                                     & 90                                                           & 92                                                                     & 0.17                                                          & 91                                                                      & 0.20                                                                \\ \hline
VGG19                                              & 94.15                                                        & 95                                                     & 94                                                           & 96                                                                     & 0.11                                                          & 94                                                                      & 0.13                                                               \\ \hline
\begin{tabular}[c]{@{}l@{}}Ensemble\\ Model\end{tabular} & 96                                                        & 97                                                     & 96                                                           & 97                                                                     & 0.03                                                          & 96                                                                      & 0.04                                                                \\ \hline
\end{tabular}}
\end{table}

\subsection{Evaluating DL Models in Brain Tumor Classification: A Comparative Analysis }
\begin{figure}[t]
    \centering
    \includegraphics[height=5cm, width=\linewidth]{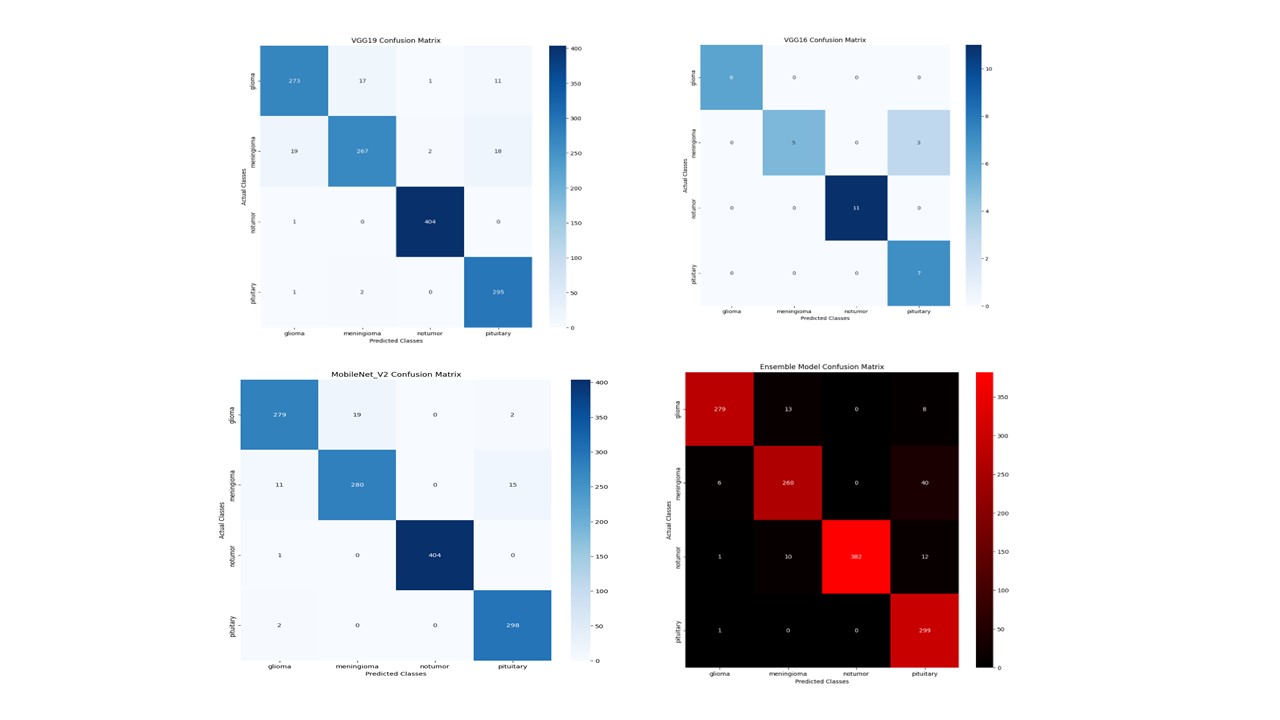}
    \caption{Ensemble \& Base models confusion matrices for Brain tumor multi-class classification )}
    \label{fig:confusion1}
    \vspace{-8mm}
\end{figure}

\begin{figure}[t]
    \centering
    \includegraphics[height=4.5cm, width=\linewidth]{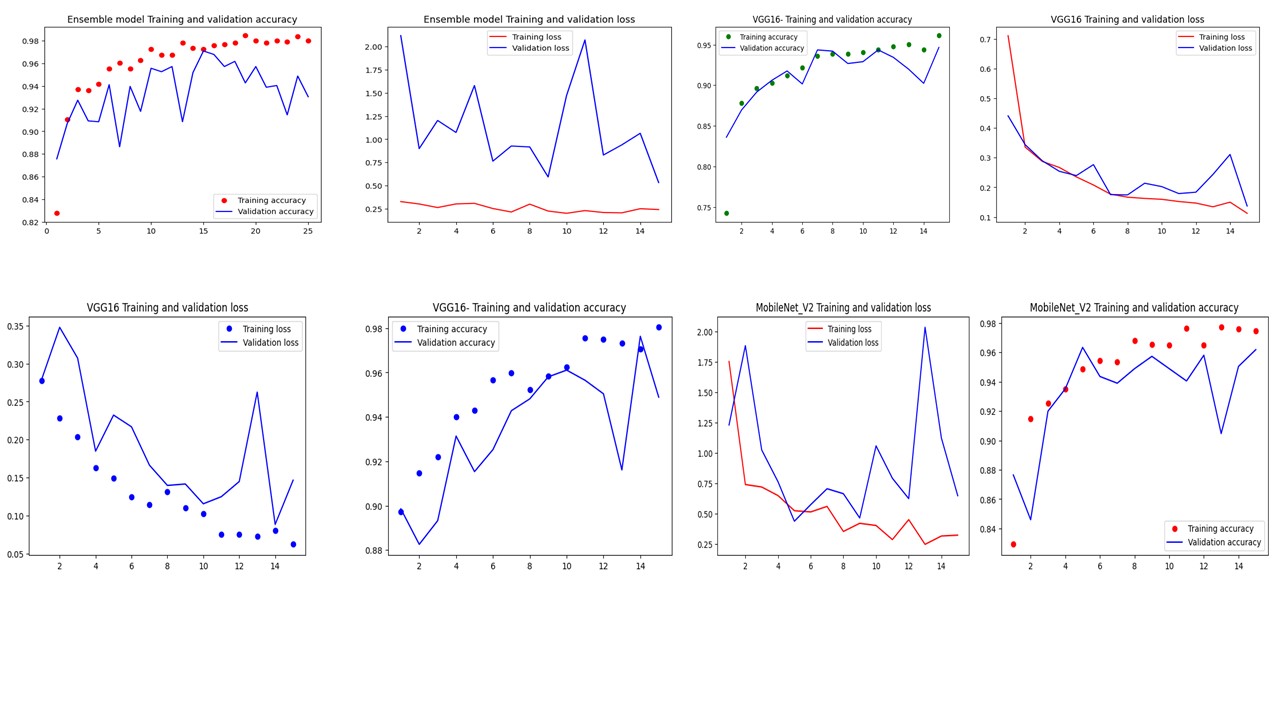}
    \caption{Ensemble \& Base models Training and Validation loss and accuracy for identifying brain tumor}
    \label{fig:loss}
    \vspace{-8mm}
\end{figure}

Our study leverages confusion matrices to assess the performance of deep learning (DL) models in brain tumor classification, including base models MobileNetV2, VGG16, VGG19, FL aggregated global model, and an ensemble model. The confusion matrix Figure \ref{fig:confusion1} and training/validation loss and accuracy metrics Figure \ref{fig:loss} are pivotal in evaluating these models. MobileNetV2 demonstrates high precision (94.19\%) and recall (92\%), with an F1-score of 94\%, indicating effective tumor classification and a balance of sensitivity and specificity. VGG16, though slightly less precise at 90.62\%, maintains a 91\% recall and an F1-score of 90\%, suggesting reliable detection capabilities with some scope for improvement in reducing training and validation losses. VGG19 shows notable recall (95\%) and precision (94.15\%), achieving an F1-score of 94\%, highlighting its sensitivity and low rate of false negatives. Notably, the ensemble model surpasses individual models with a precision of 96\%, recall of 97\%, and an F1-score of 96\%, demonstrating its superiority in reducing variance and bias and enhancing prediction accuracy for brain tumor classifications.

\subsection{Global model (FL) evaluation matrices analysis}
\begin{figure}[t]
    \centering
    \includegraphics[height=5cm, width=\linewidth]{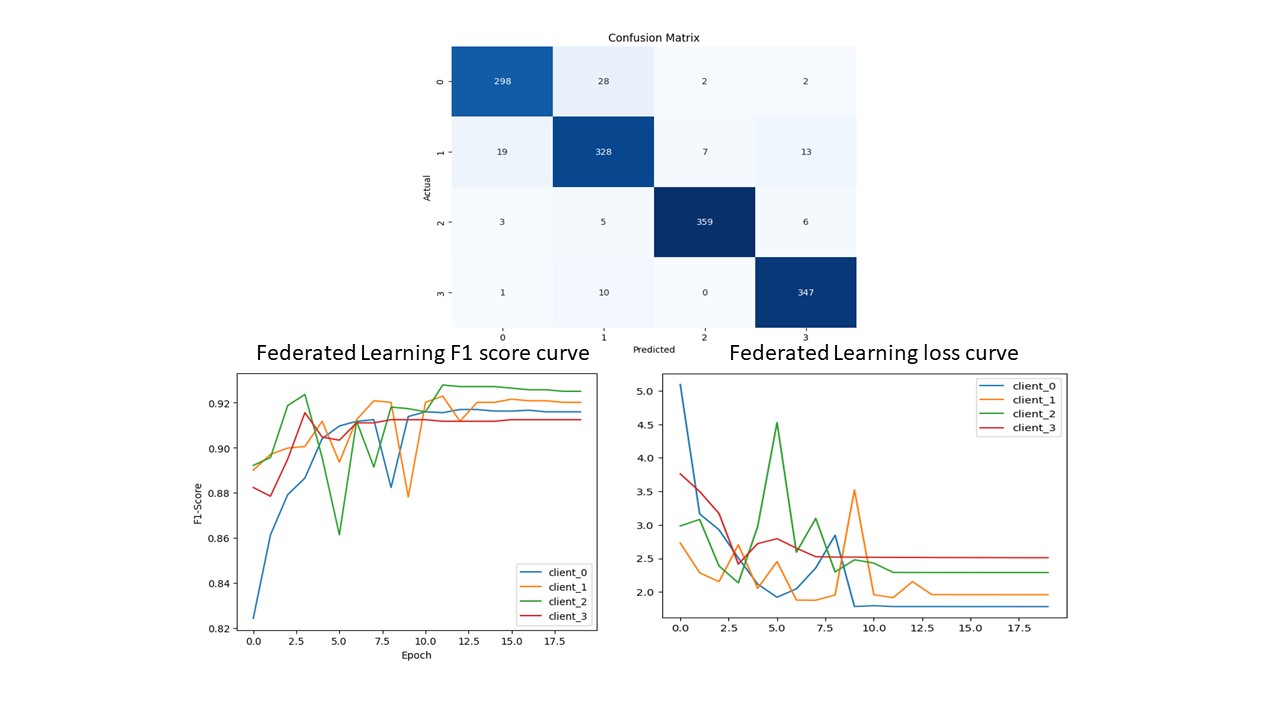}
    \caption{Federated Learning model confusion matrix for brain tumor classification at the local server )}
    \label{fig:federatedconf}
    \vspace{-4mm}
\end{figure}

The FL approach has culminated in a global model Figure \ref{fig:federatedconf} that showcases commendable precision and recall, with scores of 93.27\% and 91\%, respectively. The model achieves an F1-score of 92\%, reflecting its effective learning from distributed data sources while maintaining privacy and data security. The low training and validation losses indicate a well-fitted model with good generalization capabilities on unseen data.

This section delves into the evaluation of the global model's performance, focusing on the analysis of the F1 score and validation loss Figure \ref{fig:federatedconf} trends to gauge the efficacy and reliability of the model over the validation dataset.

\section{Conclusion}
Our study introduces the innovative EBFL framework, seamlessly integrating Federated Learning (FL) with a custom ensemble of advanced deep learning architectures. This approach significantly enhances feature extraction capabilities, surpassing single-model systems' performance while reducing bias. The EBFL framework prioritizes patient privacy by focusing on feature extraction rather than direct data sharing. Our results demonstrate the model's exceptional accuracy, with the Global model achieving a 94\% precision rate and the Ensemble model further elevating this to 96\%. However, the framework's main challenge lies in its computational complexity, presenting a valuable area for future research to explore and optimize.

 \section*{Acknowledgment}
This material is based on work supported by the
National Science Foundation Award Numbers 2205773 and 2219658.

\bibliographystyle{ieeetr}
\bibliography{References.bib}

\end{document}